\documentclass[runningheads]{llncs}
\usepackage[T1]{fontenc}
\usepackage{graphicx}
\usepackage[utf8]{inputenc} %
\usepackage[T1]{fontenc}    %
\usepackage{url}            %
\usepackage{nicefrac}       %
\usepackage{microtype}      %
\usepackage{lipsum}		%
\usepackage{doi}

\usepackage{cite}
\usepackage{amsmath, amssymb,amsfonts}
\usepackage{graphicx}
\usepackage{comment}
\usepackage{mwe}
\usepackage{hyperref}
\usepackage{epstopdf}
\usepackage{kotex}
\usepackage{tabularx}
\usepackage{url}
\usepackage{tablefootnote}

\usepackage[utf8]{inputenc}
\usepackage{algorithm}
\usepackage{algpseudocode}

\usepackage{multirow}
\usepackage{booktabs}
\usepackage{tabularx}
\usepackage{textcomp}
\newcolumntype{x}[1]{>{\centering\arraybackslash\hspace{0pt}}p{#1}}

\usepackage{adjustbox} %
\usepackage{pdflscape} %
\usepackage{lscape} %

\newcolumntype{C}{>{\centering\arraybackslash}X}

\newfloat{lstfloat}{tb}{lop}
\floatname{lstfloat}{Listing}

\usepackage{listings} %
\usepackage{seqsplit} %
\usepackage{xcolor} %

\usepackage{eucal}

\newboolean{anonymous}  %
\setboolean{anonymous}{false}  %

\makeatletter
\def\lst@lettertrue{\let\lst@ifletter\iffalse}
\makeatother

\begin{document}
\title{180 Days After EIP-4844: \\ Will Blob Sharing Solve Dilemma \\ for Small Rollups?}
\titlerunning{180 Days After EIP-4844: Blob Sharing for Rollups}

\ifthenelse{\boolean{anonymous}}
  {}  %
  {

\author{Suhyeon Lee \inst{1} \inst{2}}

\institute{
Tokamak Network\\
\email{suhyeon@tokamak.network}
\and
Korea University\\
\email{orion-alpha@korea.ac.kr}
}

  }

\maketitle              %
\begin{abstract}

The introduction of blobs through EIP-4844 has significantly reduced the Data Availability (DA) costs for rollups on Ethereum. However, due to the fixed size of blobs at 128 KB, rollups with low data throughput face a dilemma: they either use blobs inefficiently or decrease the frequency of DA submissions. Blob sharing, where multiple rollups share a single blob, has been proposed as a solution to this problem. This paper examines the effectiveness of blob sharing based on real-world data collected approximately six months after the implementation of EIP-4844. By simulating cost changes using a simple blob sharing format, we demonstrate that blob sharing can substantially improve the costs and DA service quality for small rollups, effectively resolving their dilemma. Notably, we observed cost reductions in USD exceeding 85\% for most of the rollups when they cooperate, attributable to the smoothing effect of the blob base fee achieved through blob sharing.

\end{abstract}

\section{Introduction}
\label{section: introduction}

Rollups are a fundamental layer 2 application introduced to enhance the scalability of Ethereum. While earlier solutions like Plasma aimed to increase transaction throughput, they encountered security challenges, particularly in ensuring Data Availability (DA)—the assurance that transaction data is accessible and verifiable. Rollups address these issues by providing data availability at a higher layer, thereby enhancing the overall security and reliability of the network. Rollups are primarily categorized into optimistic rollups and zk rollups, differing mainly in their approaches to achieving finality through fraud proofs and validity proofs, respectively. In the Ethereum environment, the main costs associated with rollups are DA, commitment, and proofs in case of zk rollups. Even when rollup transactions are compressed, the calldata costs remain significantly burdensome.

To alleviate the substantial DA costs that are common across all rollups, Ethereum adopted EIP-4844 \cite{EIP4844} with the Dencun upgrade on March 13, 2024. EIP-4844 introduces a new type of transaction, type-3, which includes blobs—data that are stored for approximately 4096 epochs (about 18 days). The 4096-epoch data availability period ensures that rollup validators do not encounter issues related to DA. The gas fees for blobs are significantly lower compared to calldata, thereby reducing the financial burden. On the other hand, unlike calldata, each blob is always 128 KB in size, and a single block can include up to six blobs. Due to the fixed size of blobs, users must pay gas fees corresponding to 128 KB even if they use less data.

Previous theoretical research in this area has explored various aspects of DA and transaction batching within the Ethereum ecosystem. For instance, Bar-On and Mansour \cite{bar2023optimal} and Mamageishvili and Felton \cite{mamageishvili2023efficient} analyzed efficient batching strategies related to DA submission and delays. Crapis et al. \cite{crapis2023eip} investigated strategies for transitioning between blobs and calldata, introducing the concept of blob sharing from a game-theoretic perspective. Park et al. \cite{park2024impact} conducted a comprehensive survey on the Ethereum network environment, focusing on the impact of EIP-4844.

Building on this foundation, our study concentrates on the practical implications of blob sharing, particularly highlighting the observation that real-world rollups are not utilizing blobs efficiently. As illustrated in Table \ref{table:overview}, smaller rollups frequently encounter a dilemma due to insufficient transaction data. Specifically, these rollups either cannot efficiently utilize the limited blob space (i.e., Metal, Zircuit, Boba Network) or must extend their blob submission delays to accommodate their transaction volumes (i.e., Zora, Mode, Debank Network). Given that our study analyzes approximately six months of rollup data following the implementation of EIP-4844, we consider this an appropriate time to evaluate the effectiveness of blob sharing mechanisms.

Our research question is: "Does blob sharing effectively improve the blob utilization efficiency of small rollups?" To answer this, we analyze blob fee costs and blob gaps, both before and after blob sharing with simulation, using nearly six months of Ethereum blob data.

The contributions of this paper are twofold:
\begin{itemize}
    \item Our research demonstrates the potential impact of blob sharing based on real-world data collected six months after the implementation of EIP-4844, showing that blob sharing can reduce overall costs by approximately 80\% to as much as 99\%.
    \item Our findings propose that the benefits of blob sharing extend beyond the direct rollup participants, positively influencing the overall rollup and Ethereum ecosystem.
\end{itemize}

This paper is organized as follows. Section \ref{section: model} describes the data collected and the simulation models used. Section \ref{section: results} presents the simulation results along with their interpretation. Finally, Section \ref{section: conclusion} concludes the paper with discussing future directions.

\begin{table}[ht]
\centering
\caption{Rollup Blob Utilization and Performance: Blocks 19426589 to 20611514}
\label{table:overview}
\adjustbox{max width=\linewidth}{%
\scriptsize
\begin{tabularx}{\linewidth}{l>{\centering\arraybackslash}X>{\centering\arraybackslash}X>{\centering\arraybackslash}X>{\centering\arraybackslash}X}
\toprule
\textbf{Rollup} & \textbf{Blob Count} & \textbf{Average Size (KB)} & \textbf{Total Size (GB)} & \textbf{DA Quality} \\
\midrule
Base         & 690356  & 126.51 (98.84\%)  & 83.293  & 0.309  \\
Arbitrum     & 375014  & 127.52 (99.63\%)  & 45.607  & 0.288  \\
Taiko        & 328805  & 30.70 (23.99\%)   & 9.628   & 0.592  \\
OP Mainnet   & 265469  & 126.71 (98.99\%)  & 32.080  & 0.239  \\
Scroll       & 139199  & 105.35 (82.30\%)  & 13.985  & 0.332  \\
Blast        & 106033  & 115.03 (89.87\%)  & 11.632  & 0.255  \\
Linea        & 93604   & 117.64 (91.90\%)  & 10.501  & 0.237  \\
StarkNet     & 62361   & 127.98 (99.98\%)  & 7.611   & 0.242  \\
zkSync Era   & 62149   & 128.00 (99.99\%)  & 7.586   & 0.215  \\
Paradex      & 44300   & 127.99 (99.99\%)  & 5.407   & 0.234  \\
Metal        & 31382   & 0.18 (0.14\%)     & 0.005   & 0.220  \\
Zircuit      & 27821   & 6.04 (4.72\%)     & 0.160   & 0.275  \\
Kroma        & 24193   & 59.65 (46.60\%)   & 1.376   & 0.208  \\
Zora         & 23592   & 126.58 (98.89\%)  & 2.848   & 0.161  \\
Mode         & 22585   & 126.56 (98.88\%)  & 2.726   & 0.159  \\
Rari         & 19583   & 31.08 (24.28\%)   & 0.581   & 0.214  \\
Optopia      & 7851    & 16.19 (12.65\%)   & 0.121   & 0.179  \\
Boba Network & 6236    & 3.29 (2.57\%)     & 0.020   & 0.166  \\
Debank Chain & 3358    & 113.84 (88.94\%)  & 0.365   & 0.162  \\
Camp Network & 3292    & 98.39 (76.87\%)   & 0.309   & 0.136  \\
Nal          & 2998    & 0.21 (0.17\%)     & 0.001   & 0.168  \\
Mint         & 2515    & 90.89 (71.01\%)   & 0.218   & 0.140  \\
Lambda       & 2336    & 103.80 (81.09\%)  & 0.231   & 0.143  \\
Lumio        & 1390    & 0.43 (0.33\%)     & 0.001   & 0.137  \\
Parallel     & 1348    & 119.29 (93.20\%)  & 0.153   & 0.113  \\
XGA          & 1031    & 98.20 (76.72\%)   & 0.097   & 0.142  \\
Lisk         & 811     & 8.04 (6.28\%)     & 0.006   & 0.126  \\
Kinto        & 204     & 13.20 (10.31\%)   & 0.003   & 0.124  \\
\bottomrule
\end{tabularx}
}
\end{table}

\section{Blob Sharing Model}
\label{section: model}

This section details the data utilized in the study, the evaluation criteria, and the simulation model employed.

\subsection{Collected Data and Metrics}

We analyzed data spanning from block number 19426589, starting with the implementation of the Dencun upgrade on March 13, 2024, to block number 24848485, generated around the end of August. Most of the block information and market price data were sourced from \url{Dune.com}, adhering to Dune's datasets for rollup labeling. Consequently, some rollup data remained unlabeled and were therefore excluded from the sharing simulation, despite existing as blobs. Due to minor discrepancies in the actual blob data sizes within Dune's datasets\footnote{Dune's curated dataset for blob not only stripped the right-side zero paddings from the blobs but also removed data from the left side.}, we directly calculated them by obtaining raw data from \url{blobscan.com} and removing the trailing zero padding.

We focused on three primary metrics related to blob usage:

\begin{enumerate}
    \item \textbf{Average Blob Size:} This metric indicates the amount of data a rollup actually utilizes out of the total 128 KB allocated per blob.
    \item \textbf{DA Service Quality:} Calculated using the average interval between blocks where blobs were submitted, as defined in Equation \ref{equation: da quality definition}.
    \item \textbf{Total DA cost (USD):} Calculated by summing the gas costs for calldata, priority fees, and blob usage.
\end{enumerate}

\begin{equation} \label{equation: da quality definition}
    \texttt{DA\_quality} = \frac{1}{\log(\texttt{avg\_block\_gap}) + 1}
\end{equation}

Configured this way, DA quality reaches 1 when blobs are submitted at the minimum block interval (i.e., as frequently as possible) and approaches 0 as the interval increases. A logarithmic function is used to handle large block intervals for small rollups, as taking a simple reciprocal would result in values too small for meaningful comparison. The logarithmic scaling allows for easier numerical comparisons, effectively representing DA service quality across different rollup scenarios.

\subsection{Simulation Model}

The simulation aimed to reconstruct costs based on data from all labeled rollups, incorporating up to 128 KB per blob. To achieve this, we applied several assumptions and rules. 
Blob sharing data are stored in a repeated structure. Each structure includes three components: a signature, a data length, and the actual rollup data. This design ensures proper blob submissions and facilitates efficient parsing of shared blobs.

The structure is defined as $\left( \texttt{signature} \ || \ \texttt{length} \ || \ \texttt{rollup\_data} \right)$. 
In this structure, the \texttt{signature} is 32 bytes and used to verify the rollup data, the \texttt{length} is 2 bytes and indicates the size of the rollup data, and the \texttt{rollup\_data} has a variable length, containing the actual data submitted by the rollup. Multiple such structures are concatenated together, forming a blob-sharing submission.

\textbf{Assumptions} The following assumptions were made to construct the foundational data for blob sharing:
\begin{itemize}
    \item \textbf{Constant Data Rate:} All rollups are assumed to produce data at a consistent rate prior to each blob submission. For instance, if a rollup submits a 100 KB blob at block $X+5$ following a previous submission at block $X$, we assume that the rollup generates 20 KB of data per block between blocks $X+1$ and $X+5$.
    \item \textbf{Separate DA Contract for Blob Sharing:} Blobs shared via blob sharing are submitted to a virtual blob sharing DA contract.
    \item \textbf{Uniform Gas Consumption for DA Transactions:} The gas consumption of DA transactions, including those involving blob sharing, is uniformly assumed to be 21,000 gas units. This is a minimal gas amount for DA using blobs. Some rollups (e.g., Arbitrum and Scroll) consume more gas in DA transactions for additional functionalities.
    \item \textbf{No Impact on Ethereum Base Fee and Priority Fee:} Transactions reduced by blob sharing do not affect the Ethereum base fee, which remains unchanged. Additionally, we assumed that all transactions paid the median priority fee for the block.
\end{itemize}

\textbf{Simulation Procedure} Based on these assumptions, the simulation was conducted as follows:

\begin{enumerate}
    \item \textbf{Data Preprocessing:} For each labeled rollup, the actual size and submission intervals of the blobs were used to calculate how much data each rollup produced in every block.
    \item \textbf{Blob Reconstruction:} Starting from block 19,426,589, when blobs were first submitted, the data for each labeled rollup was accumulated. Once the data reached 128 KB, it was packed into a blob for submission.
    \item \textbf{Handling Excess Data/Blob:} If rollup data exceeds 128 KB, the surplus data is configured as a new blob. And if the number of blobs generated in the current block exceeded six, the excess blobs were deferred and submitted in the next block. If six or fewer blobs were generated, they were bundled into a transaction and submitted within the current block.
    \item \textbf{Submission of Shared and unlabeled Blobs:} When submitting transactions containing blobs, if unlabeled blobs already existed in the actual data for that block, the transaction first prioritized submitting those unlabeled blobs. Any remaining space was filled with shared blobs, and any overflow was deferred to the next block.
\end{enumerate}

\section{Simulation results}
\label{section: results}

In this section, we summarize the blob sharing results and explain the greater-than-expected cost improvements through the blob fee mechanism. 

\begin{table}[htb]
\centering
\caption{Comparison of Real Data and Simulation Data (Blob Sharing)}
\label{table: results}
\adjustbox{max width=1\linewidth}{ %
\begin{tabular}{l>{\centering\arraybackslash}p{2.5cm} >{\centering\arraybackslash}p{2.5cm} >{\centering\arraybackslash}p{2.5cm}| >{\centering\arraybackslash}p{2.5cm} >{\centering\arraybackslash}p{2.5cm} >{\centering\arraybackslash}p{2.5cm}}
\toprule
\multirow{2}{*}{\textbf{Rollup}} & \multicolumn{3}{c|}{\textbf{Real Data}} & \multicolumn{3}{c}{\textbf{Simulation Data (Blob Sharing)}} \\
\cmidrule(r){2-4} \cmidrule(l){5-7}
 & \textbf{Total Cost\tablefootnote{Note that there is a difference between the actual costs of DA transactions for some rollups and the calculated values presented here. This discrepancy arises because only the minimal gas (21,000) for a Type-3 transaction is considered in the calculation. Additionally, to ensure a fair comparison with the simulation data, we assumed that transactions paid the median priority fee of their respective block.}} \newline \textbf{(USD)} & \textbf{Total Blob} \newline \textbf{Size (GB)} & \textbf{DA Service} \newline \textbf{Quality} & \textbf{Total Cost} \newline \textbf{(USD)} & \textbf{Total Blob} \newline \textbf{Size (GB)} & \textbf{DA Service} \newline \textbf{Quality} \\
\midrule
Base          & 1,720,837  & 83.293  & 0.309  & 254,949  & 83.351  & 0.980 \\
Arbitrum      & 1,037,822  & 45.607  & 0.288  & 155,725  & 45.656  & 0.983 \\
OP Mainnet    & 803,896    & 32.080  & 0.239  & 119,298  & 32.125  & 0.889 \\
Taiko         & 505,940    & 9.628   & 0.592  & 16,303   & 9.651   & 0.892 \\
StarkNet      & 328,677    & 7.611   & 0.242  & 47,764   & 7.650   & 0.818 \\
Linea         & 296,700    & 10.501  & 0.237  & 40,638   & 10.538  & 0.850 \\
Scroll        & 257,746    & 13.985  & 0.332  & 33,592   & 14.015  & 0.873 \\
zkSync Era    & 247,516    & 7.586   & 0.215  & 36,903   & 7.625   & 0.836 \\
Blast         & 172,685    & 11.632  & 0.255  & 24,924   & 11.655  & 0.991 \\
Paradex       & 98,657     & 5.407   & 0.234  & 15,101   & 5.446   & 0.834 \\
Zora          & 84,690     & 2.848   & 0.161  & 12,308   & 2.886   & 0.819 \\
Mode          & 70,086     & 2.726   & 0.159  & 10,281   & 2.764   & 0.838 \\
Metal         & 46,917     & 0.005   & 0.220  & 109.41   & 0.040   & 0.839 \\
Kroma         & 40,034     & 1.376   & 0.208  & 2,648    & 1.404   & 0.816 \\
Rari          & 33,932     & 0.581   & 0.214  & 1,146    & 0.603   & 0.836 \\
Zircuit       & 30,959     & 0.160   & 0.275  & 232.91   & 0.173   & 0.999 \\
Optopia       & 13,216     & 0.121   & 0.179  & 268.76   & 0.146   & 0.844 \\
Boba Network  & 11,073     & 0.020   & 0.166  & 107      & 0.049   & 0.819 \\
Camp Network  & 6,794      & 0.309   & 0.136  & 903      & 0.340   & 0.826 \\
Debank Chain  & 4,161      & 0.365   & 0.162  & 619      & 0.383   & 0.995 \\
Parallel      & 4,032      & 0.153   & 0.113  & 591.15   & 0.179   & 0.836 \\
Mint          & 3,799      & 0.218   & 0.140  & 488.76   & 0.242   & 0.841 \\
Lumio         & 2,045      & 0.001   & 0.137  & 51.91    & 0.024   & 0.838 \\
Lisk          & 1,548      & 0.006   & 0.126  & 70.60    & 0.032   & 0.850 \\
XGA           & 940        & 0.097   & 0.142  & 169.18   & 0.110   & 0.996 \\
Kinto         & 254        & 0.003   & 0.124  & 13.67    & 0.010   & 0.999 \\
\bottomrule
\end{tabular}
}

\end{table}

\subsection{Main Results}

\begin{figure}[htbp]
    \centering
    \begin{minipage}{\textwidth}
        \centering
        \includegraphics[width=1\textwidth]{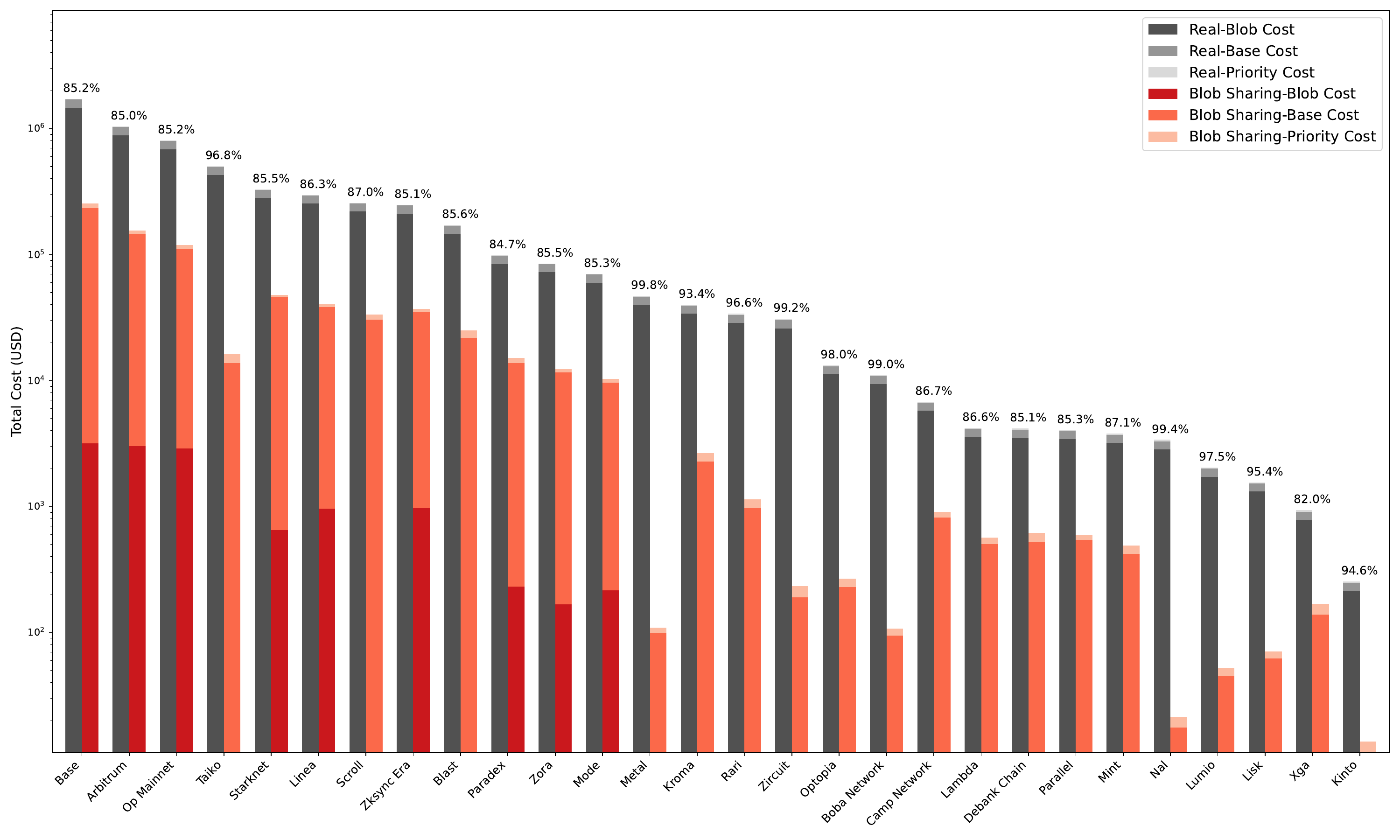}
        \caption{USD Cost Difference between Real and Simulated Blob Sharing}
        \label{fig:usd_comparison}
    \end{minipage}
    
    \vspace{0.5cm} %
    
    \begin{minipage}{\textwidth}
        \centering
        \includegraphics[width=1\textwidth]{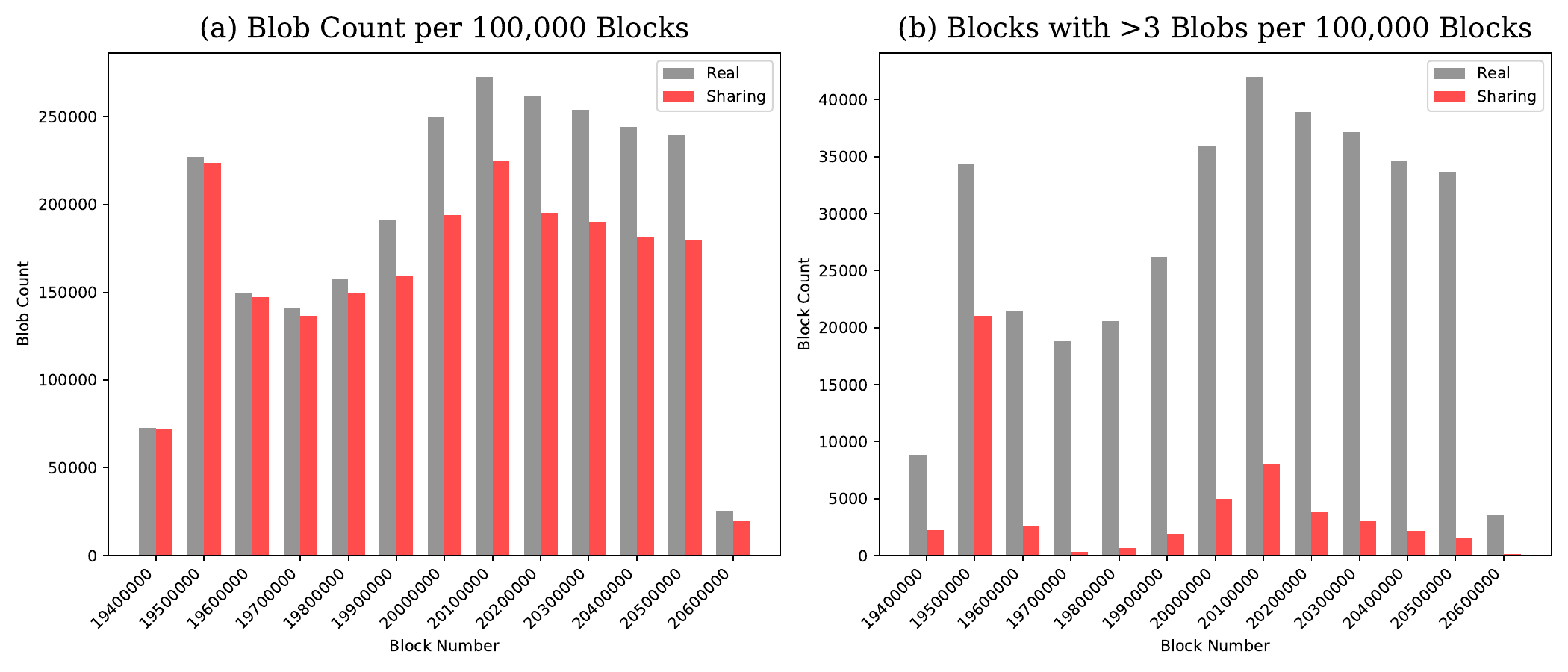}
        \caption{Blob Counts per 100,000 Blocks}
        \label{fig:number_of_blocks}
    \end{minipage}
    
    \vspace{0.5cm} %
    
    \begin{minipage}{\textwidth}
        \centering
        \includegraphics[width=1\textwidth]{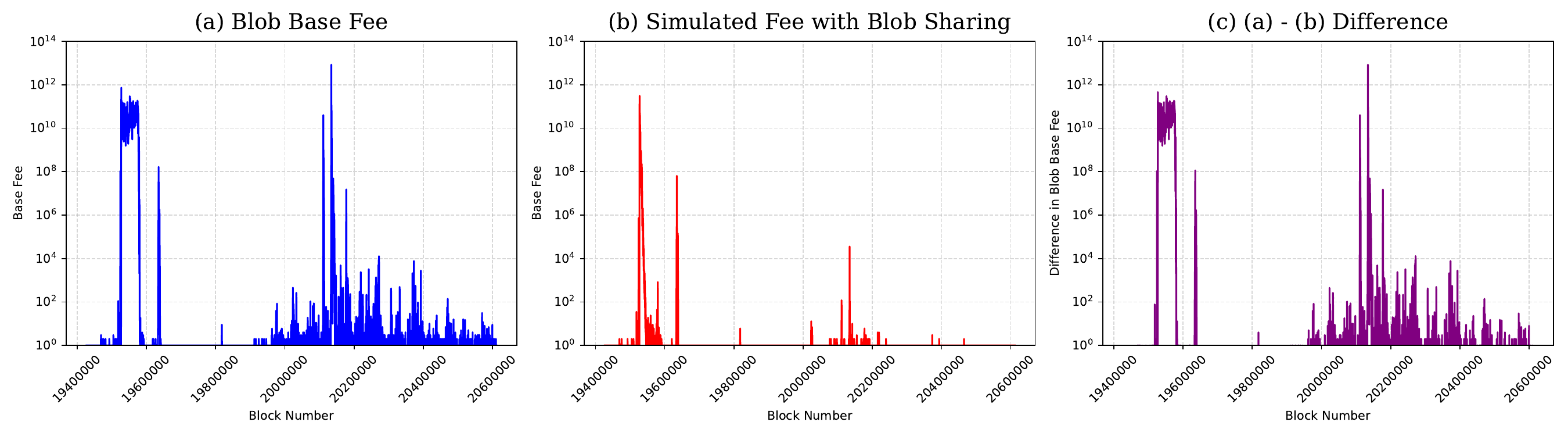}
        \caption{Blob Base Fee and Difference}
        \label{fig:fee_difference}
    \end{minipage}
\end{figure}

The simulation results are summarized in Table \ref{table: results}. Prior to the simulation, we expected cost reductions of around 50-60\% for smaller rollups. However, surprisingly, all rollups—both small and large—achieved cost reductions ranging from a minimum of 80\% to as much as 99\%. The cost differences, calculated in USD, can be seen in Figure \ref{fig:usd_comparison}, where the costs are broken down into blob, base, and priority fees. The y-axis is in logarithmic scale, explaining the high values.

Upon closer inspection, the simulation data reveals that the total blob size slightly or significantly increased compared to the real data. This is because we added the rollup signature and data length information to blobs for blob sharing, as mentioned in the previous section. As a result, smaller rollups showed a larger increase in data size, as their smaller data sets were split across blobs. Although not shown in the table, the average blob size in the simulation is 128 KB because blobs were submitted only when filled to this capacity. Additionally, the DA service quality improved for all rollups due to the frequent blob submissions enabled by blob sharing, achieving levels that even the largest rollups couldn’t reach in real-world data.

\subsection{Fee Differences}

Upon reviewing the results, we posed an additional question: Why did the overall costs decrease so significantly? We hypothesize that the primary reason for the cost reduction in DA is the \texttt{blob\_base\_fee}. The \texttt{blob\_base\_fee} increases exponentially as excess blobs (the number of blobs over three) are included in blocks. In other words, the base fee decreases exponentially as the number of blocks including excess blobs decreases.

The pricing of blobs follows an exponential function based on the number of blobs, defined by the following equations:

\begin{equation}
\texttt{new\_base\_fee} = \max \left( 
    B, 
    B \times 
    \exp\left( \frac{\texttt{excess\_blob\_gas}}{F} \right) 
\right)
\end{equation}

\begin{equation}
\texttt{excess\_blob\_gas}_i = \max \left( 0, \texttt{excess\_blob\_gas}_{i-1}  + \left(
    \texttt{blob\_count}_{i} \times G - T \right)
 \right)
\end{equation}

Here, \( B \) is the minimum base fee per unit of blob gas, \( F \) is the update fraction constant, and \( \texttt{excess\_blob\_gas}_i \) represents the excess blob gas for block \( i \), calculated as the difference between the blob gas used and the target blob gas over previous blocks. \( \texttt{blob\_count}_k \) is the number of blobs in block \( k \), \( G \) is the gas used per blob, and \( T \) is the target blob gas usage (set to 3 blobs per block).

Figure \ref{fig:number_of_blocks} shows the number of blobs over 100,000 blocks. In subfigure (a), the number of blobs slightly decreased due to blob sharing. Subfigure (b) highlights that the number of blocks with more than 3 blobs decreased significantly after blob sharing. Since the blob base fee increases exponentially, this helped smooth the fees and prevented sharp spikes.

Finally, Figure \ref{fig:fee_difference} supports our hypothesis. Subfigure (a) shows the base fee in a logarithmic scale, subfigure (b) presents the base fee under blob sharing, and subfigure (c) illustrates the difference between the two. The similarity between (a) and (c) suggests that the overall base fee after blob sharing was surprisingly lower than the observed real-world base fee.

These results offer two key insights beyond the efficiency of cost reduction through blob sharing. First, since blob sharing reduces the total number of blobs in the Ethereum network, it benefits not only the rollups that participate in blob sharing but also all rollups. Second, as a derivative effect, reducing the number of Ethereum transactions also helps lower Ethereum transaction fees and alleviates the burden on validators, ultimately benefiting the entire Ethereum ecosystem.

\section{Future works and concluding remarks} \label{section: conclusion}

In this study, we investigated the impact of blob sharing on rollups in the Ethereum ecosystem, particularly focusing on small rollups that struggle with efficient blob utilization due to low data throughput. Our simulation, based on nearly six months of real-world Ethereum blob data, demonstrated that blob sharing can significantly reduce DA costs—by approximately 80\% to 99\%—and improve DA service quality across all the labeled rollups. The primary reason for this substantial cost reduction is the smoothing of the blob base fee, as blob sharing decreases the total number of blobs submitted to the network, mitigating the exponential increase in fees associated with excess blobs.

These findings suggest that blob sharing not only benefits participating rollups by lowering their costs and enhancing service quality but also positively influences the overall Ethereum network. By reducing the total number of blobs and transactions, blob sharing alleviates the load on validators and contributes to a more efficient and scalable ecosystem.

For future work, we propose the following directions:

\begin{itemize}
    \item \textbf{Implementation of Blob Sharing Protocols:} Developing and testing practical protocols for blob sharing among rollups, including mechanisms for secure data aggregation, fair cost distribution, and coordination without compromising decentralization.
    \item \textbf{Economic Incentive Analysis:} Investigating the incentive structures required to encourage rollups to participate in blob sharing, ensuring that the benefits are strategically distributed.
\end{itemize}

\ifthenelse{\boolean{anonymous}}
  {}  %
  {
\section*{Acknowledgement}

I would like to thank Boo-Hyung Lee from Tokamak Network for discussions on the rollup blob structure and Akaki Mamageishvili from Offchain Labs for discussion on the blob sharing.
  }

\bibliographystyle{splncs04}
\bibliography{reference}

\end{document}